\documentstyle[twoside,fleqn,espcrc2,epsfig]{article}

\newcommand{\eq}{\begin{equation}}
\newcommand{\en}{\end{equation}}
\newcommand{\bea}{\begin{eqnarray}}
\newcommand{\eea}{\end{eqnarray}}

\newcommand{\bdm}{\begin{displaymath}}
\newcommand{\edm}{\end{displaymath}}

\newcommand{\tp}{\tilde{\varphi}}
\newcommand{\ba}{\begin{array}}
\newcommand{\bb}{\Box}

\newcommand{\ea}{\end{array}}

\newcommand{\ZZ}{\hbox{{\rm Z{\hbox to 3pt{\hss\rm Z}}}}}

\newcommand{\Br}{\langle}
\newcommand{\kt}{\rangle}
\newcommand{\um}{\frac12}

\newcommand{\la}{\lambda}

\newcommand{\pa}{\partial}

\newcommand{\AmS}{{\protect\the\textfont2
  A\kern-.1667em\lower.5ex\hbox{M}\kern-.125emS}}
\title{A non-trivial spectrum for the trivial  $ \la \phi^4$  theory}
\author{F. Gliozzi\address{Dipartimento di Fisica Teorica, Universit\`a
di Torino, \\
        via P. Giuria 1, 10125 Torino, Italy} }
\begin{document}
\begin{abstract}
 It is pointed out that  one-component $\phi^4 $ theory in four
 dimensions has a non-perturbative
 sector that can be studied by means of an exact duality transformation
 of its Ising limit. This duality maps it to a membrane
 model. As a consequence, the $\phi^4$-theory turns out to have, in the
 broken symmetry phase, a remarkably rich spectrum  of physical states
 corresponding to membrane excitations. A numerical study of the correlators
 between dual variables allows to evaluate the masses of the first few states.
\end{abstract}
\maketitle
\section{INTRODUCTION}
Although the four-dimensional $ \phi^4 $ theory  is known to be most
likely a trivial field theory, i.e. its continuum limit is identified with
an infrared Gaussian fixed point,
it does not mean that it is necessarily useless in describing interactions
between elementary particles (for instance Higgs particles): once the
renormalized coupling $g_r$ at zero momentum has been fixed, the triviality
property tells simply us that the ultraviolet cutoff $ \Lambda $ cannot be
pushed to arbitrary high values. For small $g_r$
the upper bound on $ \Lambda $ has the asymptotic expansion
\begin{equation}
\log(\Lambda/m_r) \leq A/g_r +B \log(g_r)+C+\cdots
\label{tb}
\end{equation}
where $m_r$ is a renormalized mass.
As a consequence, this theory cannot actually reach the continuum limit.
Nevertheless, if there is a region where the cutoff is much larger
than the physical masses entering into the game, it effectively behaves like
a continuum theory and can provide an accurate description of an interacting
system. I argue in this talk that the spectrum of this theory necessarily
 contains, in the broken symmetry phase, besides the ground particle,
 a possibly infinite tower of physical states. For sufficiently low values of
 the coupling $g_r$ the ratios among the masses of these states are universal
 constants. The above statements follow directly from an exact duality
 transformation of the $4D$ Ising model, which
 is a particular limit of a lattice regularization of the $ \phi^4 $-theory.
 According to this transformation, the broken symmetry phase of the
 $4D$ Ising model is mapped to a membrane model. It follows that
  the spectrum of the $\phi^4$-theory in such a phase contains an
  infinite tower of physical states corresponding to membrane excitations.
  If the system is sufficiently close to the critical point (i.e. $g_r$
   sufficiently small) the masses of the physical states have
  identical scaling behaviours, hence their ratios are expected to be
  universal.
Note that these properties of the spectrum of $\phi^4$ theory are not specific
of $D=4$ dimensions : The low temperature phase $T<T_c$ of the Ising model
for any $D \ge 3$ has a dual description in terms of extended objects of
dimensions $p=D-2$ which yield a (possibly) infinite set of physical states.
For instance, the  $3D$ Ising model is dual, for
$ T<T_c$, to a string (i.e. $p=1$) model  which describes  the
confining phase of the $\ZZ_2$ gauge theory. The glue-ball spectrum of
this theory has been studied only recently \cite{mi}, showing the
existence of a large set of physical states, as expected. In this work
we do a similar analysis for the $4D$ Ising model. As a result,
the first few membrane states are detected.

The remaining challenge is to understand how this wide
physical spectrum of $\phi^4$-theory at $g_r\not=0$
 can be reconciled with the single particle state of the free
Gaussian theory describing its continuum limit for $D\ge 4$. As
a possible way out we observe that the excited states of the spectrum
are not directly coupled to the local fields $\phi$'s (``order operators'';
we shall use this notion in this broad sense), but rather to the
disorder operators \cite{kc}. These can be written down as non-local and
non-polynomial expressions of the $\phi$'s (see eq.(\ref{ordi})), while
non-perturbative  proofs of triviality are based on inequalities
among vacuum expectation values of  {\sl polynomials} of the order
operators \cite{af}. Perhaps the non-perturbative part of the spectrum
could be related to the observed non trivial directions in scalar theories
with non-polynomial potentials \cite{kh}.
\section{THE MODEL}
The action of the 1-component $\phi^4$-theory on the (hyper)cubic lattice
${\cal L}=\ZZ^4 $ may be parameterized as
\eq
S_\phi=\beta\, S_{links}+\sum_{x\in{\cal L}} (\phi_x^2+\lambda(\phi_x^2-1)^2)~~,
\label{action}
\en
with
\eq
S_{links}=-\sum_{x\in{\cal L}}\sum_{\mu=1}^4\phi_x\phi_{x+\hat{\mu}}\equiv
-\sum_{links}\phi_{link}
\label{links}
\en
Where $\phi$ is a real field associated to the nodes $x$ of the lattice
and $\hat{\mu}$ denotes the unit vector in the $\mu$-direction.
Once  $m_r$ and  $g_r$ have been fixed, the renormalization group
trajectories (the ``lines of constant physics'') in the plane of the bare
parameters flow toward higher values of the bare quartic coupling $\lambda$
and terminate at the $\lambda\to\infty$ limit \cite{lw},
where the action (\ref{action})
reduces to that of the Ising model (\ref{links}) with $\phi_x\in\pm1$.
In other terms, the $4D$ Ising
model is the limit theory which saturates the the triviality bound (\ref{tb}):
at fixed $g_r$ is the best approximation to the continuum limit .
\subsection{ Duality transformation}
It is well known that the Kramers-Wannier transformation can be
extended to the Ising model defined on a  lattice
of arbitrary dimensions $D$. In particular, for $D=4 $
the Ising model with action $S_{links}$
admits a dual description in terms of a dual field $\tp_\bb\in\pm1$
associated to the plaquettes of the dual lattice
$\tilde{\cal L}= (\ZZ+\um)^4$.
The dual action is given by the sum of the contributions of the elementary
cubes of $\tilde{\cal L}$:
\eq
S_{cubes}=\sum_{cubes\in\tilde{\cal L}} \tp_{cube}~~,~~\tp_{cube}=
\prod_{\bb\in cube} \tp_\bb~~
\en
where the last product runs over the six faces of the cube.
The duality transformation states that the partition function
$Z_{Ising}(\beta)=\sum_{\phi_x}\exp(-\beta S_{links}) $ is proportional to
the partition function of the dual
description $Z_{dual}(\tilde{\beta}) =\sum_{\tp_\bb}
\exp(-\tilde\beta S_{cubes})$~, with
\eq
\sinh(2\beta)\sinh(2\tilde{\beta})=1~~.
\en
It is important to stress that this duality transformation is not a
symmetry of the model. It maps one description of the dynamical
system to another description of the same system. The last equation
shows in particular that the low temperature region of the Ising model
is mapped into the strong coupling region of the dual model.
 Such a dual description has a local $\ZZ_2$  symmetry generated by
 any arbitrary function $\eta\in\pm1$ of the links of
 $\tilde{\cal L}$, through the transformation
 \eq
 \tp_\bb\to\tp_\bb \cdot\eta_\bb~~,~~
 \eta_\bb=\prod_{links\in\bb}\eta_{link} ~~.
 \label{sym}
 \en
 This is the lattice version of the generalized gauge transform of
 an antisymmetric two-index potential   $A_{\mu\nu}\sim\tp_\bb$, i.e.
 $A_{\mu\nu}\to A_{\mu\nu}+\pa_\mu\eta_\nu-\pa_\nu\eta_\mu$
 which generates the gauge invariant three-index field strength
 $F_{\mu\nu\rho}\sim\tp_{cube}$.
 The local symmetry (\ref{sym}) implies that the
 {\sl disorder} observables \cite {kc}, i.e. the ``gauge'' invariant
 quantities of the dual description, are the vacuum expectation values of
 (products of) {\sl surface operators}  $\tp_\Sigma$  associated to any
 arbitrary, closed surface $\Sigma$ of $\tilde{\cal L}$:
 \eq
 \tp_\Sigma=\prod_{\bb\in \Sigma}\tp_\bb~~.
 \en
 In the phase corresponding to the broken $\ZZ_2$ symmetry of the
 Ising model, the correlator between surface operators
 $\Br \tp_\Sigma\,\tp_{\Sigma'}\kt_{dual}$ has a
 strong coupling expansion which can be expressed
  as the sum of weighted 3D random manifolds
 having $\Sigma$ and $\Sigma'$ as boundary. In other terms, the dual of the
 $4D$ Ising model is a membrane theory, in the same sense as the strong
 coupling expansion of any lattice gauge theory can be seen as a string
 theory, formulated as a sum of weighted random surfaces with Wilson loops
 as boundary.
 Using the duality map we can study the membrane physics directly in
 the Ising model by expressing the disorder observable in terms of the
 local field $\phi_x$: For a given
 closed, not necessarily connected surface
 $\Sigma\subset\tilde{\cal L}$ choose a $3D$ manifold
 $M\subset{\cal L}$ having $\Sigma$ as boundary: $\pa M=\Sigma$.
 Denote by $\{links\perp M\}$ the set of  links orthogonal to
 $M$. One has
 \eq
 \Br \tp_\Sigma\kt_{dual}=\Br\prod_{\{links\perp
 M\}}e^{-2\beta\phi_{link}}\kt_{Ising}
 \label{ordi}
 \en
 \subsection{ The membrane spectrum }
 In order to study the physical spectrum of the membrane one has to
 define suitable correlators between surface operators with a recipe
 very similar to the one used for the glue-ball spectrum of gauge
 models:
$i)$ choose a set of $n$ closed surfaces $\{\Sigma_1,\Sigma_2,
\cdots,\Sigma_n\}$ belonging to a given $3D$ slice ${\cal S}(x_4)$ of
the dual lattice, where $x_4$ plays  the role of  ``time''
coordinate;
$ii)$ evaluate the correlation matrix
\eq
c_{ij}(t)=\Br \sum_{x\in{\cal S}(x_4)}\tp_{\Sigma_i}
\sum_{x\in{\cal S}(x_4+t)}\tp_{\Sigma_j}\kt_{dual} ~~;
\label{cor}
\en
$iii)$ project on states of definite spin and parity.

The sum over $x$ in Eq.(\ref{cor}) is a shorthand notation to denote
translational invariant combinations of surface
operators.
For sufficiently large $t$ the eigenvalues $\la_i$ of $c(t)$ have the
asymptotic form $\la_i \sim c_i e^{-m_it}$~,
where $m_i$ denotes the mass of the $i^{th}$ eigenstate.

 \section{NUMERICAL RESULTS}
 Taking advantage of eq.(\ref{ordi}), we have evaluated the masses of the
 lowest membrane states  by  studying the $4D$ Ising model with a
 non-local cluster updating algorithm \cite{sw} at $\beta=0.154$ on a
 lattice of size $12^3\times 16$. Here, in a large scale numerical
 simulation \cite{uw}, it was observed small finite volume effects and
 a good agreement with the 3-loop $\beta$-function.

 We made three
 different numerical experiments for a total of $ 3\times10^6$
 iterations, using correlators of disorder operators associated to closed
 surfaces with the topology of the sphere
 and of the torus. We chose 12 different shapes with area ranging from
 38 to 134 plaquettes and analyzed the channels $J^P=0^+,1^-,2^+$.
 In the channel $J^P=1^-$ the signal was too low to extract a reliable
 quantitative information. Two states have
 been detected in the channel $J^P=0^+$ and one in the channel $J^P=2^+$.
 Their  masses, in lattice units, are given by
 \eq
 m_{0^+}=0.5522(35)~~,
 \en
 \eq
 m'_{0^+}=1.34(10)~~~~,
 \en
 \eq
 m_{2^+}=1.74(22)~~~~.
 \en
As a consistency check, note that
the mass of the lowest membrane state $m_{0^+}$ should coincide with the
physical mass $m$ of the ground particle of the $\phi^4$-theory, because
the mass gap of the theory cannot depend on the kind of description employed
(the Ising model or its dual). Actually in ref.\cite{uw} from analysis of the
order-order correlator on a lattice of the same size the value
$m=0.553(3)$ has been reported .

 The fact that the lowest particle can be described as a membrane state
 implies that it couples both to the order operator $\phi_x$ and to the
 disorder one $\tp_\Sigma$. On the contrary the excited membrane
 states are expected to be decoupled from $\phi_x$, otherwise
 they should  manifest themselves by unitarity also in the
 order-order correlators, where no signal of this sort has been
 previously reported. Also this prediction is well supported by
 our analysis.
\vskip .3 cm


\begin{thebibliography}{90}
\bibitem{kc}L.P.Kadanoff and H.Ceva, Phys. Rev. { B3} (1971) 3918.
\bibitem{af}M.Aizenman, Phys. Rev. Lett. { 47} (1981) 1;
 J.Fr\"olich, Nucl. Phys. { B200} [FS4] (1982) 281.
\bibitem{mi}V.Agostini, G.Carlino, M.Caselle, M.Hasen\-busch,
Nucl. Phys. { B 484} (1997) 331; see also the talk of M.Caselle at
this conference.
\bibitem{kh} K. Halpern and K.Huang, Phys. Rev. Lett. {74} (1995) 3526
  and Phys.Rev. {D53} (1996) 3252.
\bibitem{lw}M.L\"uscher and P.Weisz,  Nucl. Phys. {B290} (1987) 25
and { B295} (1988) 65.
\bibitem{sw} R.H.Swendsen and J.S.Wang, Phys. Rev. Lett. {58} (1987) 86 .
\bibitem{uw} K.Jansen, I.Montvay, G.M\"unster, T.Trap\-penberg and U.Wolff,
Nucl. Phys. { B 322} (1989) 698.
\end{thebibliography}
\end{document}